\documentstyle[prd,aps,floats,tabularx,epsfig]{revtex}

\newcommand{\ETAL}{{et al.\ }}

\begin{document}

\draft
\preprint{\
\begin{tabular}{rr}
&
\end{tabular}
}
\twocolumn[\hsize\textwidth\columnwidth\hsize\csname@twocolumnfalse\endcsname
\title{Scale of Homogeneity of the Universe from WMAP}
\author{Patr\'{\i}cia G. Castro, Marian Douspis, Pedro G. Ferreira}
\address{Astrophysics, University of Oxford,
Denys Wilkinson Building, Keble Road, Oxford OX1 3RH, UK}

\maketitle

\begin{abstract}
We review the physics of the Grishchuck-Zel'dovich effect which
describes the impact of large amplitude, super-horizon
gravitational field fluctuations on the Cosmic Microwave
Background anisotropy power spectrum. Using the latest
determination of the spectrum by WMAP, we infer a lower limit on
the present length-scale of such fluctuations of $3.9\times10^3$ times  the
cosmological particle horizon (at the 95\% confidence level).
\end{abstract}
\date{\today}
\pacs{PACS Numbers : }
\renewcommand{\thefootnote}{\arabic{footnote}} \setcounter{footnote}{0}
]
\noindent A definitive map of large scale fluctuations in the Cosmic
Microwave Background (CMB) is currently being rendered by the WMAP
satellite \cite{wmapwebsite}. The preliminary results have been used to
constrain a host of cosmological parameters within the context of
a Big Bang universe with primordial adiabatic perturbations
\cite{SpergelEtAl03,VerdeEtAl03}. A candidate model, which consists of a cocktail
of cold dark, matter, baryons, neutrinos, radiation and a cosmological
constant (the $\Lambda$CDM model), seems to fit the data well,
albeit with a larger goodness
of fit than would be desired. The paradigm of an essentially homogeneous
universe with mild deviations is systematically passing many of the
observational tests with which it is confronted.

The homogeneity of space-time is one of the fundamental assumptions
of our current working model of the universe. One possibility is that
our observable universe
is in a particularly quiet patch of an inhomogeneous space time which has
undergone a period of inflation \cite{Guth81,Albrecht:Steinhardt82,Linde82}.
Conceivably there may be
indirect evidence of the primordial inhomogeneity within which our homogeneous
patch resides. Grishchuck and
Zel'dovich (GZ hereafter) have pointed out that such large scale inhomogeneity
will have an impact on the amplitude of the quadrupole of the CMB~
\cite{Grishchuk:Zeldovich77}. In this report we combine their
formalism with a goodness of fit analysis of the WMAP data to find
a lower bound on the scale of inhomogeneity of the universe.


On large scales, the Sachs-Wolfe (SW) effect~\cite{Sachs:Wolfe67}
relates metric perturbations with the
CMB temperature anisotropies. It is given, along a direction
$\hat{n}$ in the sky, in terms of the perturbations in the
gravitational potential, $\Phi$, by
\begin{eqnarray}
\frac{dT}{T}(\hat{n}) = \frac{1}{3}\; &&\Phi(\hat{n}(\eta_0 -
\eta_{LS}),\eta_{LS}) \nonumber \\
                       && + \;2 \int_{\eta_{\small{LS}}}^{\eta_0} d\eta
                         \frac{d\Phi}{d\eta} (\hat{n}(\eta - \eta_{LS}),\eta)
\label{eq:dT_T_sachs_wolfe}
\end{eqnarray}
where $\eta$ is the conformal time defined today as $\eta_0$.
The conformal
time today is the comoving distance to the horizon.
The second term is the Integrated Sachs-Wolfe (ISW)
effect and accounts for the contributions to the fluctuations due to
the possible variations of the gravitational potential along the line
of sight whereas the first term is due to perturbations in the
potential present at the last scattering (LS) surface.

By expanding the temperature fluctuations in spherical harmonics on the sky,
it is straightforward to show that, in a flat Universe,
\begin{eqnarray}
C_{\ell} =  \frac{2}{\pi}\;
        &&\int_{0}^{+\infty} k^2dk P_{\Phi}^{LS}(k)
           [ \frac{1}{3}j_{\ell}(k(\eta_0-\eta_{LS})) \nonumber \\
        &&+\; 2 \int_{\eta_{LS}}^{\eta_0} d\eta \frac{dg(\eta)}{d\eta}
                                        j_{\ell}(k(\eta_0-\eta)) ]^2
\label{eq:cl_sachs-wolfe}
\end{eqnarray}
where we have used $\Phi({\mathbf x},\eta)=\int \Phi({\mathbf
k},\eta) e^{-i{\mathbf x}.{\mathbf k}}d^3k/(2\pi)^3$ and ${\mathbf
k}$ is the comoving wave-number related to $x$ by $k = 2\pi / x$.
For the suppression factor $g$, defined as $\Phi({\mathbf
k},\eta)=g(\eta)\Phi({\mathbf k},\eta_{LS})$, we use the precise
parametrization of~\cite{Carroll92}. Note that $g(\eta_{LS})=1$
and $g$ varies with time depending on the cosmological model
considered. $P_{\Phi}^{LS}$ is the gravitational field power
spectrum at the last scattering surface. With a flat spectrum
$P_{\Phi}^{LS}(k) = A k^{-3}$, scales much larger than the horizon
make little contribution to the $C_{\ell}$'s: If we consider only
the SW term in Eq.~(\ref{eq:cl_sachs-wolfe}), the wave-number
contributions such that $k \ll 1/\eta_{0}$ scale with $k^{2l}$.
But a possible sharp and localized increase in the power spectrum
at a certain super-horizon scale $k_{GZ}$ of the type
$P_{\Phi}^{LS}(k) = A \delta (k-k_{GZ})$ would leave a stronger
imprint in the $C_{\ell}$'s at low multipoles as firstly shown by
Grishchuck and Zel'dovich in \cite{Grishchuk:Zeldovich77}. This
effect is more prominent for the monopole, dipole and quadrupole
components. The monopole cannot be distinguished from a small
shift in the measured temperature of the CMB because the density
wave is super-horizon sized and we are averaging over our horizon.
The dipole is contaminated by the Doppler effect caused by the
motion of our Local Group with respect to the CMB rest-frame. The
relevant multipole to study the GZ effect is therefore the
quadrupole. For this multipole, contributions from super-horizon
gravitational inhomogeneous modes of characteristic scale $k$ to
the GZ effect get suppressed by a factor of ${\cal O}(k^4)$, as will
be shown later. By not seeing an increase of power in the
quadrupole, one can therefore place constrains on the homogeneity
properties of super-horizon perturbations namely on the minimum
allowed wavelength. On larger scales, due to the
suppression factor of ${\cal O}(k^4)$, the universe can be
anisotropic and inhomogeneous without being in disagreement with
the available experimental constraints~\cite{Grishchuk92}. In
fact, if we were to consider contributions from larger scales as
well, (assuming, for example, a step function $P_{\Phi}^{LS}(k) = A \theta
(k_{GZ}-k)$), the total contribution to the quadrupole value would
decrease, doubling the value obtained for $k_{GZ}$ from the data.
This behaviour is due to the way the normalization constant $A$ is
derived. As will be explained below, in this case, when fixing
$A$, we would also be considering contributions from modes smaller
than $k_{GZ}$ which would reduce the relative amplitude of the
potential fluctuations coming solely from the $k_{GZ}$ mode, as
compared to the standard Grishchuck and Zel'dovich calculation.

We consider that the homogeneity of the gravitational field
fluctuations is satisfied as long as the characteristic amplitudes
of the fluctuations are below one (in the linear regime boundary).
In the limiting case of amplitudes of order $1$, which we will
investigate, the amplitude $A$ of the power spectrum can be
obtained by imposing that $<|\Phi(L_{GZ})|^2> = 1$, where
$<|\Phi(L_{GZ})|^2>$ is the average variance of the gravitational
field fluctuations in spheres of comoving radius $L_{GZ}/2$ given
by $L_{GZ}/2 = 2\pi/k_{GZ}$. We point out that there is some
confusion in the literature in this
respect~\cite{Kashlinsky94,Turner91}: we are fixing the amplitude
of the gravitational field perturbation associated with the
density perturbation, not the amplitude of the density
perturbation.

The normalized expression for the Grishchuck-Zel'dovich effect is the
following
\begin{eqnarray}
C_{\ell} = \frac{2^6 \pi^5}{9}\;&&
           [  \frac{1}{3}j_{\ell}(k_{GZ}(\eta_0-\eta_{LS})) \nonumber \\
           && +\; 2 \int_{\eta_{LS}}^{\eta_0} d\eta \frac{dg(\eta)}{\eta}
                                            j_{\ell}(k_{GZ}(\eta_0-\eta))]^2
\label{eq:cl_GZ}
\end{eqnarray}
At $l=2$,
for $k_{GZ} \simeq 10^{-6}$ Mpc$^{-1}$ and for the cosmologies considered here,
the effect of the ISW corresponds to around 10 \% of the pure SW contributions.
Looking at the purely SW term of this expression, which dominates
over the ISW term, we can use the small-argument
limit of the spherical Bessel function
$j_{2}(x) = x^{2}/15$, to obtain the following relation
for the quadrupole amplitude
\begin{equation}
 \frac{\Delta T}{T} (\ell=2) \simeq (k_{GZ} \eta_{0})^2 =
                                    \left( 2 \pi \frac{L_{0}}{L_{GZ}} \right)^2
\label{eq:GZ_quadrupole}
\end{equation}
where
$\Delta T / T = \sqrt{\ell(\ell+1)C_{\ell}/(2\pi)}$,
$L_{0}=2\eta_0$ is
the present size of the horizon diameter and $L_{GZ}$ is the present
length-scale of the super-horizon homogeneous patch of characteristic
wave-number $k_{GZ}$. This approximation was obtained previously by others
~\cite{Grishchuk92}
without the $2\pi$ factor which was neglected.
Having an upper limit on the value of the
quadrupole, we can obtain a lower limit on the diameter size of the largest
possible homogeneous scale of the universe. Similar expressions for
the GZ effect were calculated for the open cosmology
case~\cite{Kashlinsky94} for which the constraints on $L_{GZ}$ can be
even more stringent. Given the present experimental consensus
around the flatness of our universe, we place ourselves in
a flat geometry.

\begin{figure}[!t]
\begin{center}
\resizebox{\hsize}{!}{\includegraphics[angle=0,totalheight=7cm,width=8.cm]
{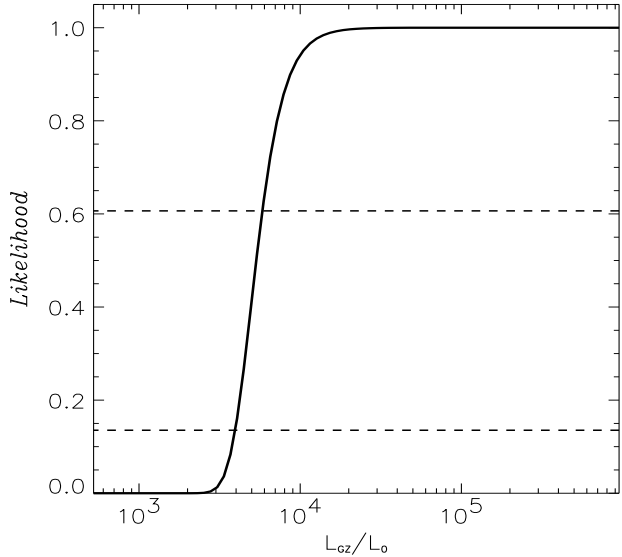}}
\end{center}
\caption{Likelihood of the ratio $L_{GZ}/L_{0}$. $L_{GZ}$
is the length-scale of super-horizon gravitational field fluctuations
of amplitude of order $1$ inferred by the WMAP temperature
data. $L_0$ is the horizon scale. All the other cosmological
parameters investigated have
been maximized over. The lower horizontal line
shows the 95\% and the higher one, the 68\% confidence
level on $L_{GZ}/L_{0}$. } \label{fig}
\end{figure}

We proceed by using the recent WMAP temperature data to constrain the
possible contribution of super-horizon gravitational field
fluctuations (with an amplitude of order $1$) to the quadrupole.
For the analysis we consider the
GZ contribution to the power spectrum in addition to the usual power
law $\Lambda$--CDM spectrum. We compute the standard
$C_\ell$ spectrum by using CAMB code~\cite{cambwebsite}, 
and add to it the GZ spectrum computed
analytically as described in Eq.~(\ref{eq:cl_GZ}). We consider a grid of
models corresponding to inflationary adiabatic perturbations in a flat
cosmology by varying the following cosmological parameters around the
best values found by the WMAP collaboration \cite{SpergelEtAl03}:
$H_0,\;\Omega_\Lambda,\;\tau,\;n$ and $L_{GZ}/L_0$. The
normalization of the ``power law'' $C_\ell$
spectrum was left free and was marginalized over. No contributions from
neutrinos or gravitational waves were considered. As the effect of the
super-horizon perturbation is expected at very low $\ell$ (mainly at
the quadrupole), where the presence of the baryons is negligible,
we fixed the baryon contribution at the best estimate
from WMAP: $\Omega_bh^2=0.023$. We used the likelihood code provided
by the WMAP team \cite{VerdeEtAl03} and modified by \cite{Lewis} to
compute the likelihood of each model of our grid. We then maximized
over all the parameters, except $L_{GZ}/L_0$, in order to retrieve the
likelihood as a function of the scale of the super-horizon
perturbations. The result is plotted in Fig.~\ref{fig}.

We obtain an upper limit on the region of homogeneity (and isotropy)
of $L_{GZ} > 3927 L_{0}$ (at 95\% CL). This correspond to
$L_{GZ} \sim 6.9  \times 10^7 \;
\rm Mpc$ for $H_0=68$ Kms$^{-1}$Mpc$^{-1}$.
As expected this value is comparable (except
for the $2\pi$ factor, as detailed above) to previous constraints
relying on the COBE
data~\cite{Kashlinsky94,Turner91,StoegerEtAl97}, which also
presented a low quadrupole value.
The estimate from
\cite{Kogut96} gives $Q_{COBE}=10\pm 3 \pm 7 \; \mu K$, where the
quoted errors are statistical and systematic
(from the foreground removal) respectively.
The WMAP results~\cite{BennetEtAl03}
give roughly the same upper limit of $Q_{WMAP}=12.3\pm3.1 \; \mu K$
at 68\% CL, where the systematics were assumed to be
negligible. As stressed before, the contribution from the GZ effect
to the quadrupole is proportional to $k_{GZ}^4$ such that
observing a quadrupole amplitude
higher by a factor of $2$ decreases the length-scale $L_{GZ}$ by only a
factor of $2^{1/4}$. The small difference in the upper limit for the
quadrupole between the WMAP and the COBE experiments
induces an even smaller difference on
the determination of the possible scale of the gravitational
fluctuations considered. Nevertheless our result is more
robust because we include the present uncertainty on
most of the relevant cosmological parameters.
It is worth mentioning that -- as could have been predicted --
nearly no degeneracies were found between the scale of the
super-horizon fluctuations and all the other
parameters explored in this analysis. This is due to
the unique and strong localization of the GZ effect
at the quadrupole component.

There has been some debate surrounding the statistical significance
of the low quadrupole values obtained by the WMAP
and the COBE DMR experiments as compared to
the values predicted by the standard cosmological
model~\cite{SpergelEtAl03,BennetEtAl03,TegmarkEtAl03,GaztanagaEtAl03,Efstathiou03a,deOliveira-CostaEtAl03}.
In this context, some work has been devoted to finding a physical
mechanism able to explain the possible discrepancy
~\cite{SpergelEtAl03,GaztanagaEtAl03,TegmarkEtAl03,KawasakiTakahashi03,FengZhang03,ClineEtAl03,ContaldiEtAl03,BridleEtAl03a,Efstathiou03b,BridleEtAl03b,NiarchouEtAl03}.
In this report, we take the quadrupole and octopole
results from WMAP at face value and assume
they are in accordance with the standard model within the
cosmic variance errors, regardless of the present controversy.
Of course, if any of these proposed mechanisms,
modifying the standard scenario, are indeed
lowering the quadrupole value
than our constraints
on the homogeneity scale would be slightly affected.


To conclude we point out that one can estimate the number of
e-foldings, $N_{e}$, which a patch of the universe has undergone
during an inflationary period. Numerical studies of the onset of
inflation in inhomogeneous universes indicate that the region of
space time which inflates must start of with at most fluctuations
of order one on the horizon scale \cite{GoldwirthPiran92,MatznerEtAl93}.
Mapping this scale onto $L_{GZ}$ today we find that $N_e\simeq
\ln[M_{inf}/(10^{-15}\;{\rm GeV})]$ where $M_{inf}$ is the energy scale
of inflation. For inflation occurring at $M_{inf}>10^{16}\;{\rm GeV}$
one has
$N_{e}>72$. It is interesting (but not surprising) that $N_{e}$
is similar to the number of e-folding required to get the correct
amplitude and slope of fluctuations from the parametric
amplification of quantum fluctuations during inflation
\cite{HuiDodelson03,LiddleLeach03}. One can also
reinterpret this result as a constraint on the duration of
an {\it uninterrupted} period of inflation and in doing
so limit the effects of primeval, non-thermal corrections to
cosmological perturbations~\cite{KaloperKaplinghat03}.

{\it Acknowledgments}: PGC was supported by the Funda\c{c}\~{a}o
para a Ci\^{e}ncia e a Tecnologia under the reference PRAXIS
XXI/BD/21249/99. M.D. is supported by a EU CMBNet fellowship. PGF
thanks the Royal Society. We thank an anonymous referee whose comments
helped in improving the manuscript.


\tighten


\vspace{-.3in}
\begin{thebibliography}{99}
\vspace{-.6in}
\bibitem{wmapwebsite} See the webpage of WMAP: http://map.gsfc.nasa.gov/
\bibitem{SpergelEtAl03} D. N. Spergel \ETAL (WMAP collab.) (2003), {\tt [astro-ph/0302209]}
\bibitem{VerdeEtAl03} L. Verde \ETAL (WMAP collab.) (2003), {\tt [astro-ph/0302218]}
\bibitem{Guth81} A. Guth, Phys. Rev. D 23, 347 (1981)
\bibitem{Albrecht:Steinhardt82} A. Albrecht and P. J. Steinhardt,
Phys. Rev. D Lett. 48, 1220 (1982)
\bibitem{Linde82} A. Linde, Phys. Lett. B108, 389 (1982)
\bibitem{Grishchuk:Zeldovich77} L. P. Grishchuk and Ya. B. Zel'dovich,
Astron. Zh. 55, 209, (1978) [Sov. Astron. 22, 125 (1978)]
\bibitem{Sachs:Wolfe67} R. K. Sachs and A. M. Wolfe, Astrophys.
J. 147, 73 (1967)
\bibitem{Grishchuk92} L. P. Grishchuk, Phys. Rev. D 45, 4717 (1992)
\bibitem{Carroll92} S. M. Carroll, W. H. Press and E. L. Turner,
Ann. Rev. Astron. Astrophys. 30, 499 (1992)
\bibitem{cambwebsite} See the webpage of CAMB: http://camb.info/
\bibitem{Lewis} A. Lewis, MCMC: http://cosmologist.info/cosmomc/
\bibitem{Kashlinsky94} A. Kashlinsky, I. I. Tkachev and J. Frieman,
Phys. Rev. Lett. 73, 1582 (1994)
\bibitem{Turner91} M. S. Turner, Phys. Rev. D 44, 3737 (1991)
\bibitem{StoegerEtAl97} W. R. Stoeger,  M. E. Araujo and T. Gebbie, Astrophys.
J., 476, 435 (1997)
\bibitem{Kogut96}  A. Kogut, A. J. Banday, C. L. Bennett,
K. M. Gorski, G. Hinshaw, G. F. Smoot and E. I. Wright,
Astrophys. J. Lett. 464, L5 (1996)
\bibitem{BennetEtAl03} C. L. Bennett \ETAL (WMAP Collab.) (2003), {\tt [astro-ph/0302207]}
\bibitem{TegmarkEtAl03} M. Tegmark, A. de Oliveira-Costa and A. Hamilton (2003),
{\tt [astro-ph/0302496]}
\bibitem{GaztanagaEtAl03} E. Gaztanaga \ETAL (2003), {\tt [astro-ph/0304178]}
\bibitem{Efstathiou03a} G. Efstathiou (2003), {\tt [astro-ph/0306431]}
\bibitem{deOliveira-CostaEtAl03} A. de Oliveira-Costa \ETAL (2003), {\tt [astro-ph/0307282]}
\bibitem{KawasakiTakahashi03} M. Kawasaki and F. Takahashi (2003),
{\tt [astro-ph/0305319]}
\bibitem{FengZhang03} B. Feng and X. Zhang (2003),
{\tt [astro-ph/0305020]}
\bibitem{ClineEtAl03} J. M. Cline, P. Crotty and J. Lesgourgues (2003),
{\tt [astro-ph/0304558]}
\bibitem{ContaldiEtAl03} C. R. Contaldi, M. Peloso, L. Kofman, A. Linde (2003),
{\tt [astro-ph/0303636]}
\bibitem{BridleEtAl03a}  S. L. Bridle, O. Lahav, J. P. Ostriker and
P. J. Steinhardt (2003), {\tt [astro-ph/0303180]}
\bibitem{Efstathiou03b} G. Efstathiou (2003), {\tt [astro-ph/0303127]}
\bibitem{BridleEtAl03b} S. L. Bridle, A. M. Lewis, J. Weller and G. Efstathiou (2003),
{\tt [astro-ph/0302306]}
\bibitem{NiarchouEtAl03} A. Niarchou, A. H. Jaffe and L. Pogosian (2003),
{\tt [astro-ph/0308461]}
\bibitem{GoldwirthPiran92} D. Goldwirth and T. Piran, Phys. Rept.
214, 223 (1992)
\bibitem{MatznerEtAl93} H. Kurki-Suonio, P. Laguna, R. A. Matzner, Phys.
Rev. D. 48, 3611 (1993)
\bibitem{HuiDodelson03} S. Dodelson and L. Hui (2003), {\tt
[astro-ph/0305113]}
\bibitem{LiddleLeach03} A. Liddle and S. Leach (2003), {\tt
[astro-ph/0305263]}
\bibitem{KaloperKaplinghat03} N. Kaloper and M. Kaplinghat (2003), {\tt [hep-th/0307016]}
\end{thebibliography}
\end{document}